\documentclass[aps,prd,twocolumn,superscriptaddress,preprintnumbers,floatfix,nofootinbib,notitlepage,showkeys,showpacs]{revtex4-1}

\usepackage[utf8]{inputenc}

\usepackage{graphicx}
\usepackage{hyperref}
\usepackage{latexsym}
\usepackage{amsmath}
\usepackage{amssymb}
\usepackage{bbm}
\usepackage{ulem}
\usepackage{pdfsync}
\usepackage{epsfig}
\usepackage{epstopdf}
\usepackage{subfigure}
\usepackage{xcolor}
\usepackage{comment}
\usepackage{slashed}

\long\def\exclude#1{}

\newcommand{\beq}{\begin{equation}}
\newcommand{\eeq}{\end{equation}}

\allowdisplaybreaks

\bibpunct{[}{]}{,}{n}{}{,}

\newcommand{\dif}{{\rm d}}

\begin{document}

\title{Seeking dark matter with $\gamma$-ray attenuation}
\author{Jos\'e Luis Bernal}
\affiliation{William H. Miller III Department of Physics and Astronomy, Johns Hopkins University,  3400 North Charles Street, Baltimore, MD 21218, United States}
\author{Andrea Caputo}
\affiliation{School of Physics and Astronomy, Tel-Aviv University, Tel-Aviv 69978, Israel}\affiliation{ Department of Particle Physics and Astrophysics,
Weizmann Institute of Science, Rehovot 7610001,Israel}
\author{Gabriela Sato-Polito}
\affiliation{William H. Miller III Department of Physics and Astronomy, Johns Hopkins University,  3400 North Charles Street, Baltimore, MD 21218, United States}
\author{Jordan Mirocha}
\affiliation{McGill University, Department of Physics \& McGill Space Institute, 3600 Rue University, Montr\'eal, QC, H3A 2T8}
\author{Marc Kamionkowski}
\affiliation{William H. Miller III Department of Physics and Astronomy, Johns Hopkins University,  3400 North Charles Street, Baltimore, MD 21218, United States}

\date{\today}

%==========================

\begin{abstract}

The flux of high-energy astrophysical $\gamma$ rays is attenuated by the production of electron-positron pairs from scattering off of extragalactic background light (EBL).  We use the most up-to-date information on galaxy populations to compute their contributions to the pair-production optical depth.  We find that the optical depth inferred from $\gamma$-ray measurements exceeds that expected from galaxies at the $\sim2\sigma$ level.  If the excess is modeled as a frequency-independent re-scaling of the standard contribution to the EBL from galaxies, then an excess (an overall $14-30\%$ increase of the EBL) over the null hypothesis of no excess at the $2.7\sigma$ level.  If the frequency dependence of the excess is instead modeled as a two-photon decay of a dark-matter axion, then the excess is favored over the null hypothesis at the $2.1\sigma$ confidence level.  While we find no evidence for a dark-matter signal, the analysis sets the strongest current bounds on the photon-axion coupling over the $8-25$ eV mass range. This work highlights the sensitivity of $\gamma$-ray optical depth measurements to ALPs, which is expected to improve with new observatories and better EBL determinations from future observations.

\end{abstract}

\maketitle

%%%%%%%%%%%%%%
%%%%%%%%%%%%%%
\section{Introduction}
The extragalactic background light (EBL) is defined as the integrated flux aggregating all emission over cosmic times~\cite{Cooray_review}. Direct measurements are challenging due to the overwhelming contamination from foregrounds, which makes it necessary to resort to indirect observational or theoretical determinations to get a census of the populations of emitters. In addition to astrophysical emissions, exotic contributions may be present, arising from a potential connection between dark matter and Standard Model particles, for instance.

Dark matter is a key component of our understanding of the Universe, but a microscopic model backed by observational and experimental evidence is yet to be found~\cite{Bertone_DMreview}. The axion---a pseudo-Nambu-Goldstone boson initially proposed to solve the strong-CP problem---and axion-like particles (ALPs) are natural dark-matter candidates~\cite{Abbott:1982af, Dine:1982ah,Preskill:1982cy, Weinberg:1977ma, Wilczek:1977pj, Peccei:1977hh, Peccei:1977ur}, with rich phenomenology that enables a variety of search strategies~\cite{Marsh_review,axionDM_snowmass}. The coupling between ALPs and photons causes oscillations in the presence of magnetic fields, but also allows ALPs to undergo a monochromatic decay into two photons with energy $m_a c^2/2$, where $m_a$ is the ALP mass. The decay rate  depends on the ALP-photon effective coupling $g_{a\gamma}$ as $\Gamma_a = (m_a c^2)^3g_{a\gamma}^2/32h$, where $c$ is the speed of light and $h$ is the Planck constant. Searches for signatures of ALP decays span a vast energy range (see e.g.,~\cite{Caputo:2018ljp, Caputo:2018vmy,Gong_IR,Caputo_IR,Caputo:2019djj, Boyarsky:2014ska, Riemer-Sorensen:2014yda, Dessert:2018qih,Cadamuro_axion,Foster_XMM,GeweringPeine_Xraydec,  Blanco_gammaray,Cohen_gammaray,NathMaity_Pevdm,Esmaili_pevdm,Ellis_gravitino, Wadekar_LeoTDMheating, Hu_DMSpecD, Chluba_specD,Bolliet_SD,Iocco_BBN,Pospelov_BBN,Poulin_BBNDM,Poulin_decDM,Slatyer_dark,Slatyer_decDM,Lucca_synergy}). Here we seek their contributions to the EBL, focusing on indirect EBL determinations from $\gamma$-ray attenuation.

As $\gamma$ rays propagate through a bath of low-energy photons they are absorbed through electron-positron pair production~\cite{1966Natur.211..472J,Gould:1967zzb}. Joint analyses of the observed blazar spectra allow the attenuation to be determined as function of source redshift and observed $\gamma$-ray energy~\cite{FermiLAT_2012}. Assuming there is no secondary production of $\gamma$ rays due to undeflected cosmic rays in the jet~\cite{Essey:2009ju, Essey:2009zg, Essey:2010er}, these measurements can be used to infer the EBL intensity between the  infrared and near ultraviolet as a function of redshift~\cite{Finke_EBLgray,Fermilat_EBL18,Acciari_EBL,Desai_EBL}. Moreover, using independent determinations of the EBL, they can also be used to constrain the expansion history of the Universe~\cite{Dominguez_H0}, study Pop~III stars~\cite{2012MNRAS.420..800G}, and test fundamental physics~\cite{Bitau_review}, among others. In particular, $\gamma$-ray attenuation measurements are sensitive to axion-photon oscillations: using the EBL as a ``screen'' for a ``light through a wall'' experiment~\cite{Hooper:2007bq, Mirizzi_ALPconversion,Hochmuth_ALPconversion,deAngelis_ALPconversion} yields competitive bounds on the ALP-photon coupling for $10^{-2}-10^2$ neV masses (see e.g.,~\cite{HESS_ALP,Fermilat_ALP,Li_ALP,Jacobsen_ALP,Li:2021zms,Li:2022jgi}). Nonetheless, $\gamma$-ray attenuation is also sensitive to multi-electronvolt ALP decays through their contribution to the EBL, as proposed in Refs.~\cite{Kalashev_gray_ALP,Korochkin_bump,Korochkin_axion}.  

Revisiting exotic contributions to the EBL is timely. A recent measurement of the cosmic optical background performed by the New Horizons spacecraft yields a $\sim~4\sigma$ significant excess with respect to the expected flux from deep galaxy counts~\cite{NH21,NH22}. Furthermore, indirect determinations of the EBL~\cite{Fermilat_EBL18, Desai_EBL} lie above the inferred EBL from galaxy populations over a much wider frequency range~\cite{SaldanaLopez_EBL}. These excesses may be due to unaccounted for astrophysical contributions, but the New Horizons's excess  may also be explained with ALP dark matter~\cite{Bernal_COB}. 

Electronvolt-scale ALPs are hard to probe with current strategies, which motivates the development of new probes. Forthcoming line-intensity mapping experiments are expected to significantly improve sensitivities in this region of the parameter space~\cite{Creque-Sarbinowski:2018ebl,Bernal_limdm, Shirasaki_limlensdm}. Contributing to this endeavor, we present an independent probe for the dark-matter ALP using high-energy $\gamma$ rays. We significantly improve over preliminary attempts~\cite{Korochkin_bump,Korochkin_axion}, performing a systematic analysis of $\gamma$-ray absorption of a considerably larger sample of sources across redshift and observed energies, and confront it with a comprehensive observational and theoretical determination of the astrophysical EBL. 

This paper is structured as follows. We review the computation of the $\gamma$-ray optical depth and introduce the contributions considered in this work in Sec.~\ref{sec:opticaldepth}; we describe our analysis and present our results in Sec.~\ref{sec:results}; and conclude in Sec.~\ref{sec:conclusions}. Further details on the astrophysical model for the standard contributions to the EBL considered in this work, the likelihood used in the analysis, and null tests to inform our conclusions can be found in the Appendix.  

\section{$\gamma$-ray optical depth}
\label{sec:opticaldepth}
The optical depth is the line-of-sight integral of the inverse of the mean free path $l$ of all the interactions that a $\gamma$ ray encounters as it travels from the source to us: $\tau = c\int_0^{z_{\rm s}}\dif z l^{-1}/\left[H(1+z)\right]$, where $z_{\rm s}$ is the source redshift, and $H$ is the Hubble parameter. We refer to observed energies and rest-frame energies at a given redshift as $E$ and $\epsilon\equiv E(1+z)$, respectively. 
For a $\gamma$ ray with observed energy $E_\gamma$ in a bath of EBL photons, electron-positron pair production occurs above the EBL photon energy threshold,
\begin{equation}
    \epsilon_{\rm min} = \frac{2m_e^2c^4}{E_\gamma(1+z)(1-\mu)}\,,
\end{equation}
where $m_e$ is the electron mass and $\mu$ is the cosine of the angle of incidence between the two photons. The cross section for the process is~\cite{Gould:1967zzb}
\begin{equation}
\begin{split}
    \sigma_{\gamma\gamma}(\beta) & =\frac{3\sigma_{\rm T}}{16}\left(1-\beta^2\right)\times\\
    & \times\left[2\beta\left(\beta^2-2\right)+\left(3-\beta^4\right)\ln \left(\frac{1+\beta}{1-\beta}\right)\right]\,,
\end{split}
\end{equation}
where $\sigma_T$ is the Thomson cross-section and $\beta^2=1-2m_e^2c^4/(\epsilon\epsilon_\gamma(1-\mu))$ is the electron velocity in the center-of-mass frame. Assuming an isotropic gas of low-energy background photons with specific number density $\dif n/\dif \epsilon$ in proper coordinates, the inverse of the mean free path is 
\begin{equation}
    l^{-1} = \int_{0}^{\infty}\dif \epsilon\frac{\dif n}{\dif \epsilon}\int_{-1}^{1}\dif \mu \frac{(1-\mu)}{2}\sigma_{\gamma\gamma}\Theta(\epsilon-\epsilon_{\rm min})\,,
    \label{eq:mfp}
\end{equation}
where $\Theta$ is the Heaviside function.\footnote{We further simplify this equation following Ref.~\cite{Biteau_2015}, limiting the energy integral to values above $\epsilon_{\rm min}'$ and computing the angular integral analytically, changing variable from $\mu$ to $\beta$.} The angular integral peaks as $\epsilon$ tends to $\epsilon'_{\rm min}$, which corresponds to  $\epsilon_{\rm min}$ with $\mu=-1$. The number density can also be obtained from the specific intensity per unit energy $I_\epsilon$ in comoving coordinates as $\dif n/\dif \epsilon=4\pi (1+z)^3 I_\epsilon/(c\epsilon)$.\footnote{Note that $\lambda I_\lambda = \epsilon I_\epsilon$.} As a reference, the minimum energy that ever contributes to the attenuation is $\epsilon_{\rm min}'\simeq 1.3\,{\rm eV}\left(100\,{\rm GeV}/E_\gamma\right)\left(2/(1+z_{\rm s})\right)$.

\subsection{Contributions to the EBL}
The proper ALP-sourced photon number density per energy interval at a given redshift is the aggregate of all the decays at earlier times that redshift to the energy of interest:
\begin{equation}
\begin{split}
    & \left(\frac{\dif n}{\dif \epsilon}\right)^{\rm dec}  = \frac{\Omega_{a}\rho_{\rm c}c^2\Gamma_{a}(1+z)^3}{m_a c^2/2}\times \\ 
    & \qquad \times \int_z^\infty\frac{\dif z'}{H(z')(1+z')}\delta_D\left(\epsilon - \frac{m_a c^2(1+z)}{2(1+z')}\right) = \\ 
    & = \frac{2\Omega_{a}\rho_{\rm c}c^2\Gamma_a(1+z)^3}{m_a c^2\epsilon H(z_*)}\Theta(z_\star-z)\,,
\end{split}
\label{eq:dnde_dm}
\end{equation}
where $\delta_D$ is the Dirac delta function, $\Omega_a$ and $\rho_{\rm c}$ are the ALP density parameter and the critical density today, respectively, and $z_*\equiv m_a c^2(1+z)/(2\epsilon)-1$ is the redshift of decay that contributes to the photon energy and redshift of interest. We assume that ALPs make up all of the dark matter and, for the energies of interest, we can safely consider that decay photons have a Dirac delta-function profile~\cite{Gong_IR}. The condition $z_\star > z$ is equivalent to setting the rest-frame energy as the upper limit in the integral over $\epsilon$ in Eq.~\eqref{eq:mfp}. 

In order to quantitatively determine the contribution from ALP decays to the $\gamma$-ray attenuation we need to budget standard contributions to the EBL from astrophysical sources. The EBL can be directly measured (see e.g., Refs.~\cite{Mattila_17,Matsuura_CIBER,Matsuoka_pioneer,Matsumoto_2018}), though foreground contamination hinders this approach (motivating measurements from the outskirts of the solar system~\cite{Zemcov_NHCOB,NH21,NH22}). Instead, we reconstruct the EBL using observational and theoretical information. We distinguish three components: emission from galaxies at $z<6$, emission from galaxies at $z>6$, and the intra-halo light (IHL)---emission from a faint population of stars tidally removed from galaxies~\cite{Cooray_2012, Zemcov_2014, Matsumoto_19}. The cosmic microwave background photons have energies too low to contribute to the optical depth of the blazars of interest. 

We use the determination of the EBL sourced from galaxies at $z<6$ from Ref.~\cite{SaldanaLopez_EBL}, which is fairly consistent with alternative semianalytic, phenomenological and empirical studies (see e.g., Refs.~\cite{Finke_EBL10,Dominguez_EBL11,Hegalson_EBL12,Gilmore_EBL12,Driver_EBL16, Stecker:2016fsg}). We reweight specific comoving luminosity densities $j_\nu(z)$ inferred from multiwavelength HST/CANDELS surveys~\cite{Grogin_candels,Koekemoer_candels} to match our fiducial cosmology.\footnote{Full \textit{Planck} data set best-fit parameters assuming $\Lambda$CDM~\cite{Planck:2018vyg}.} We use the code 
\textsc{ARES}\footnote{\url{https://github.com/mirochaj/ares}}~\cite{Mirocha:2014faa} to model the contribution from stars at $z>6$, using an empirical model for Pop II star-forming galaxies calibrated with observations of the ultraviolet luminosity function~\cite{Mirocha2017}. We also consider the contribution from Pop III stars, following the modeling from Ref.~\cite{Mirocha:2017xxz}. In each case, we also include X-ray emission representative of high-mass X-ray binary systems, though these sources contribute little to the $\gamma$-ray optical depth. We adopt extreme cases  to be conservative regarding uncertainties. Finally, we model the IHL luminosity density with a power-law dependence in mass and redshift~\cite{Cooray_2012}, assuming a spectral energy distribution similar to old elliptical galaxies~\cite{Krick:2007fp}. We take results from Ref.~\cite{Mitchell-Wynne:2015rha} to set the fiducial contribution and its uncertainties. We compute the flux from each astrophysical contribution $i$ in comoving coordinates as function of redshift,
\begin{equation}
    \lambda I_{\lambda,i}(\lambda,z) = \frac{c^2}{4\pi\lambda}\int_{z_{0}}^{z_{\rm max}}\dif z'\frac{j_{\nu',i}\left(\frac{\lambda(1+z)}{1+z'},z'\right)}{H(z')(1+z')}\,,
    \label{eq:lamIlam}
\end{equation}
where $\lambda$ is the rest-frame wavelength at redshift $z$, $\nu'$ is the rest-frame frequency at $z'$, and $z_0=z\,(6)$ and $z_{\rm max}=6\,(60)$ for the contributions from galaxies at $z<6$ and the IHL (from galaxies at $z>6$). We use the resulting flux as function of redshift to compute the optical depth due to each contribution, propagating the uncertainties accordingly. The interested reader can find more details about the computation of the astrophysical contributions in App.~\ref{sec:AppA}.

\begin{figure}[t]
 \begin{centering}
\includegraphics[width=\columnwidth]{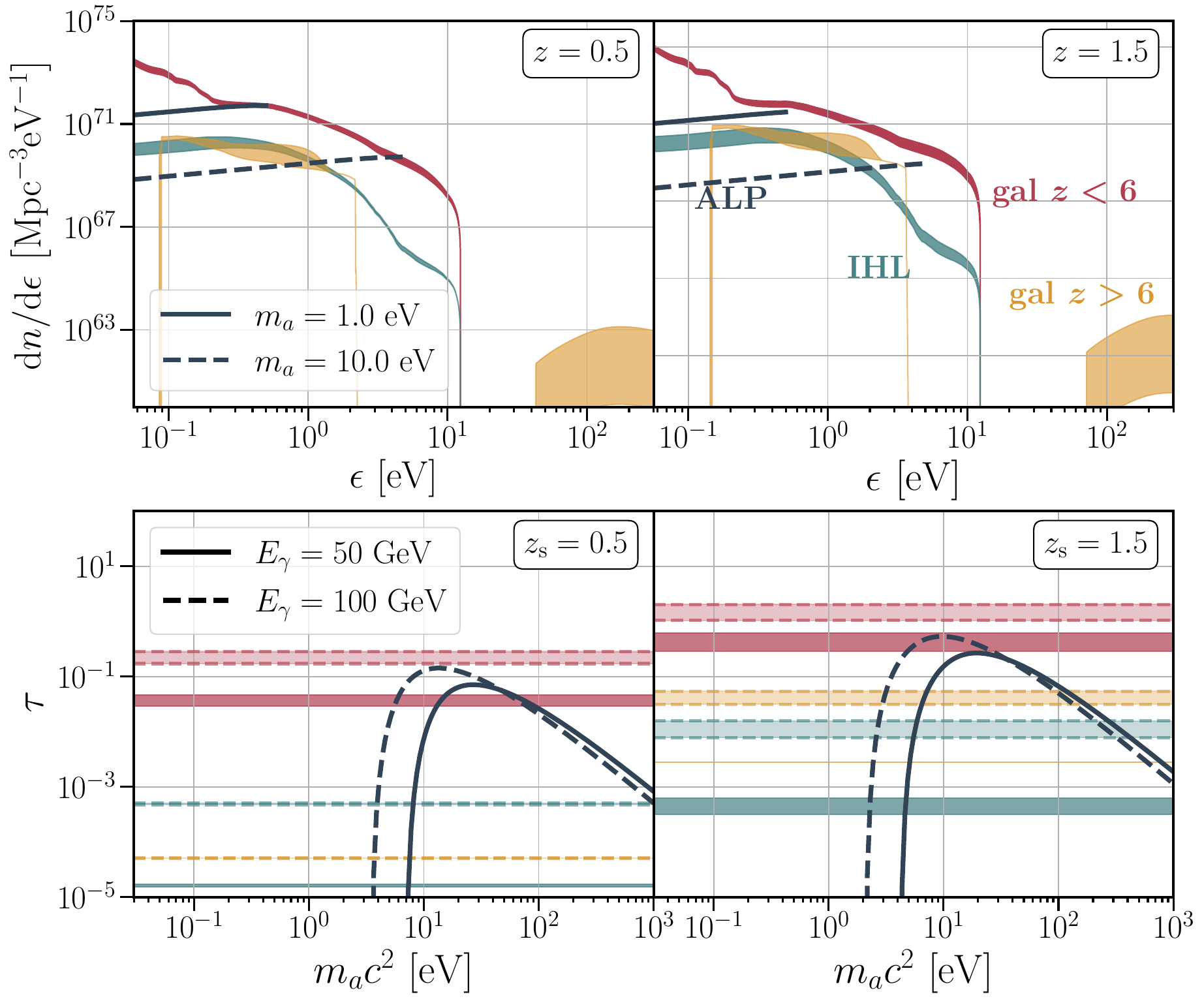}
\caption{Photon number density in proper coordinates (top panels) and $\gamma$-ray optical depth for  observed energies of 50 and 100 GeV (bottom panels), distinguishing contributions from galaxies at $z<6$, $z>6$, and the IHL (red, orange and blue, respectively, shaded bands denoting the 68\% confidence level uncertainties) and ALP decays for a decay rate of $3\times 10^{-24}\,{\rm s^{-1}}$ (solid lines), similar to the best fit obtained in this work.}  
\label{fig:example}
\end{centering}
\end{figure}

In order to provide an idea of the shape and magnitude of each contribution, we compare the various photon number densities at $z=0.5$ and $1.5$, and their corresponding  optical depths for $\gamma$ rays with observed energies of 50 and 100 GeV in Fig.~\ref{fig:example}. The ALP-sourced photon density grows with $\epsilon$, contrary to astrophysical photons, since for the same ALP mass lower energies correspond to decays occurring at higher redshifts, for which the time interval, and hence the total flux, is smaller. The dependence on redshift is mostly a small change in amplitude, with an additional small tilt and shift in the astrophysical contributions. As expected, the contribution from galaxies at $z<6$ vastly dominates the EBL. The gap between low and high-energy EBL photons for astrophysical sources at $z>6$ is due to neutral hydrogen absorption. The dependence of the optical depth on source redshift is similar, with a slight shift towards smaller masses for higher source redshifts. 

\section{Analysis and results}
\label{sec:results}
We can confront now the predicted optical depth due to ALP decays with attenuation measurements. We take public optical depth measurements obtained from the analysis of 739 blazars observed by FermiLAT~\cite{Fermilat_EBL18} and 38 blazars observed by ground-based Cherenkov observatories~\cite{Desai_EBL}. These amount to optical depth measurements in 12 $z_{\rm s}$ bins for $E_{\gamma}\in\left[10-10^3\right]$ GeV, and in 2 $z_{\rm s}$ bins for $E_{\gamma}\in\left[0.1-20\right]$ TeV. We combine these measurements with the inferred optical depth and uncertainty from astrophysical sources to generate a sample of residual optical depth $\tau_{\rm res}$ to be contributed from unaccounted-for sources (i.e., the ALP decays in this case). 
We remove lower-limit measurements from our sample since they do not set any limit to the additional contributions. Further details on the computation of $\tau_{\rm res}$ can be found in App.~\ref{sec:AppB}  
We find $\tau_{\rm res}>0$ in some cases, which is consistent with previous results comparing EBL determinations with the inference from $\gamma$-ray attenuation~\cite{SaldanaLopez_EBL}.

We follow Ref.~\cite{Fermilat_EBL18} and assume a Gaussian likelihood for $\tau_{\rm res}$. We adopt flat priors on $\log_{10}(m_ac^2/{\rm eV})\in [-1,2]$ and $\log_{10}(\Gamma_a/{\rm s^{-1}})\in [-28,-20]$. We evaluate the unnormalized posterior in a fine grid in $\log_{10}(m_ac^2/{\rm eV})$ and $\log_{10}(\Gamma_a/{\rm s^{-1}})$, and derive the posterior from the $\Delta\chi^2$ with respect to the best-fit $\chi^2$.\footnote{Exploring the parameter space of $\log_{\rm 10}(\Gamma_a/{\rm s^{-1}})$ and $\log_{\rm 10}(g_{a\gamma}/{\rm GeV^{-1}})$ returns the same posterior, since the likelihood $\mathcal{L}(\log_{\rm 10}(g_{a\gamma}/{\rm GeV^{-1}}),\log_{10}(m_ac^2/{\rm eV})) = 2\mathcal{L}(\log_{\rm 10}(\Gamma_a/{\rm s^{-1}}),\log_{10}(m_ac^2/{\rm eV}))$, which yields the same $\Delta\chi^2$.} We neglect neutral hydrogen absorption for ALP masses above 20.4 eV. The absorption of the photons produced in these decays would require higher decay rates to contribute the same to the EBL. However, as shown below, this region of the parameter space is better constrained by other probes, so that this choice does not affect our conclusions.

\begin{figure*}[t]
 \begin{centering}
\includegraphics[width=\textwidth]{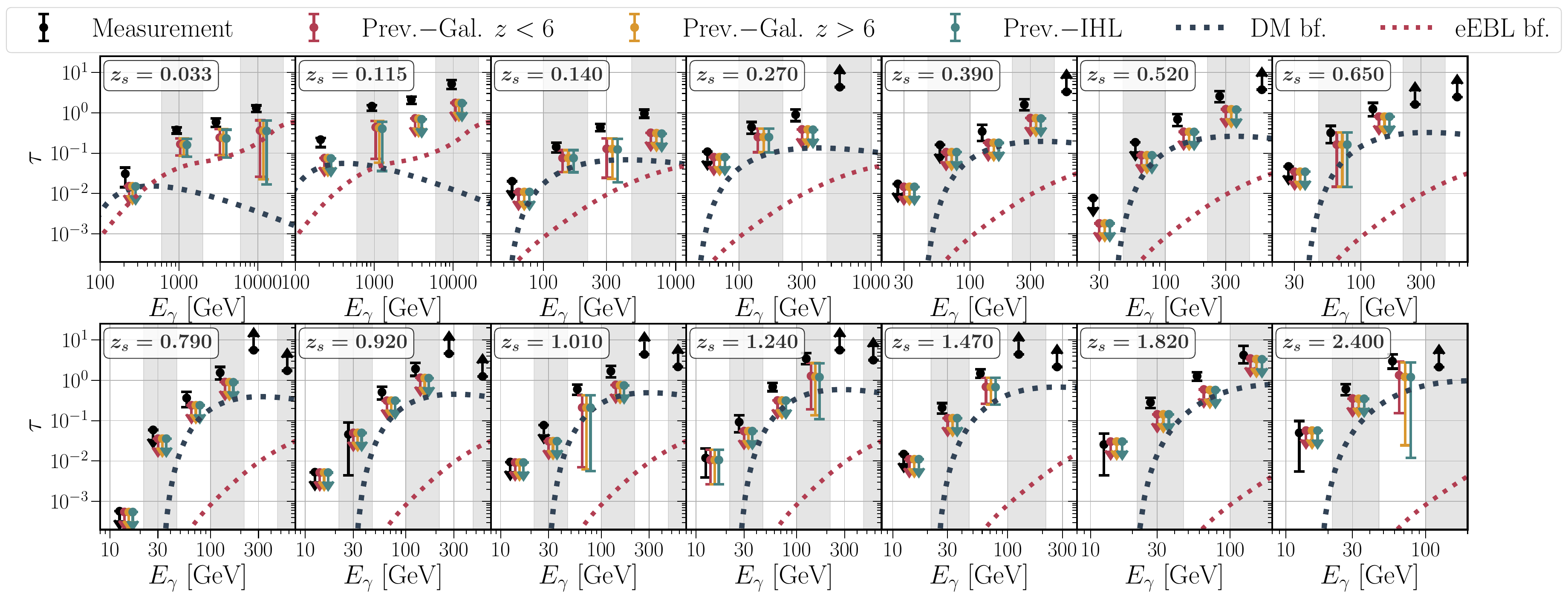}
\caption{Measured and residual $\gamma$-ray optical depths binned in source redshift and observed $\gamma$-ray energy. We show residual optical depths as we cumulatively subtract astrophysical components (as indicated in the legend, each entry is the previous minus the indicated component). Dark blue dotted lines show the best-fit results for the contributions from ALP decays. We also show the best fit results for the case with additional extragalactic background with red dotted lines.}  
\label{fig:chi2}
\end{centering}
\end{figure*}

The excess in optical depth can be explained with ALP decays, with  
best fit $m_ac^2=9.1\,{\rm eV}$ and $\Gamma_a=2.5\times 10^{-24}\,{\rm s^{-1}}$, which  corresponds to $g_{a\gamma}=2.1\times 10^{-11}\,{\rm GeV^{-1}}$. We find that the contributions from ALP decays are preferred over the null hypothesis (no additional optical depth) at $2.1\sigma$ significance ($p-{\rm value}=0.038$). We show $\tau_{\rm meas}$, $\tau_{\rm res}$ and the predicted contributions from ALPs for the best fit in Fig.~\ref{fig:chi2}. The ALP contributions provide a very good fit for $\tau_{\rm res}$, with the exception of the measurements at $z_{\rm s}=0.033$. 

Our results can also easily be rescaled to smaller abundances of ALPs modifying Eq.~\eqref{eq:dnde_dm} accordingly. For the coupling and masses of interest, the correct dark-matter relic abundance can be obtained via the misalignment mechanism~\cite{Preskill:1982cy, Abbott:1982af,Dine:1982ah}, although only in non-standard cosmology scenarios~\cite{Blinov:2019rhb, Arias:2012az,Eroncel:2022vjg} or specific photo-philic models where the ALP-photon coupling is enhanced, while keeping a relatively large decay constant~\cite{Farina:2016tgd}.  

We also explore a simplified null test. We consider a separate case to explain the optical depth excess with a boosted contribution from galaxies at $z<6$: we multiply such contribution by a constant factor $(1+F_{\rm eEBL})$,\footnote{Note that we only increase the mean contributions and do not vary the errorbars that go into the computation of $\tau_{\rm res}$.} adopting a flat prior on $\log_{10}F_{\rm eEBL}\in \left[-3,1\right]$. We find  $F_{\rm eEBL}=0.22\pm 0.08$ at 68\% confidence level, which fits well the first bins in $z_{\rm s}$, but fails for blazars at higher redshift for which $\tau_{\rm res}$ is not an upper limit. $F_{\rm eEBL}>0$ is preferred at $2.7\sigma$ ($p-{\rm value}=6.7\times 10^{-3}$).  The $\Delta\chi^2$ between the best fits in this and the ALP cases favors the former with marginal significance. If the first $z_{\rm s}$ bin is removed we find $\Delta\chi^2=-2.9$, favoring ALP decays, with approximately the same best-fit parameters for the ALP decay than before (and same significance of detection against the null hypothesis), but $0.17\pm 0.09$ for $F_{\rm eEBL}$ at 68\% confidence level. Further discussion and null tests can be found in App.~\ref{sec:AppC}. 

\begin{figure}[t]
 \begin{centering}
\includegraphics[width=\columnwidth]{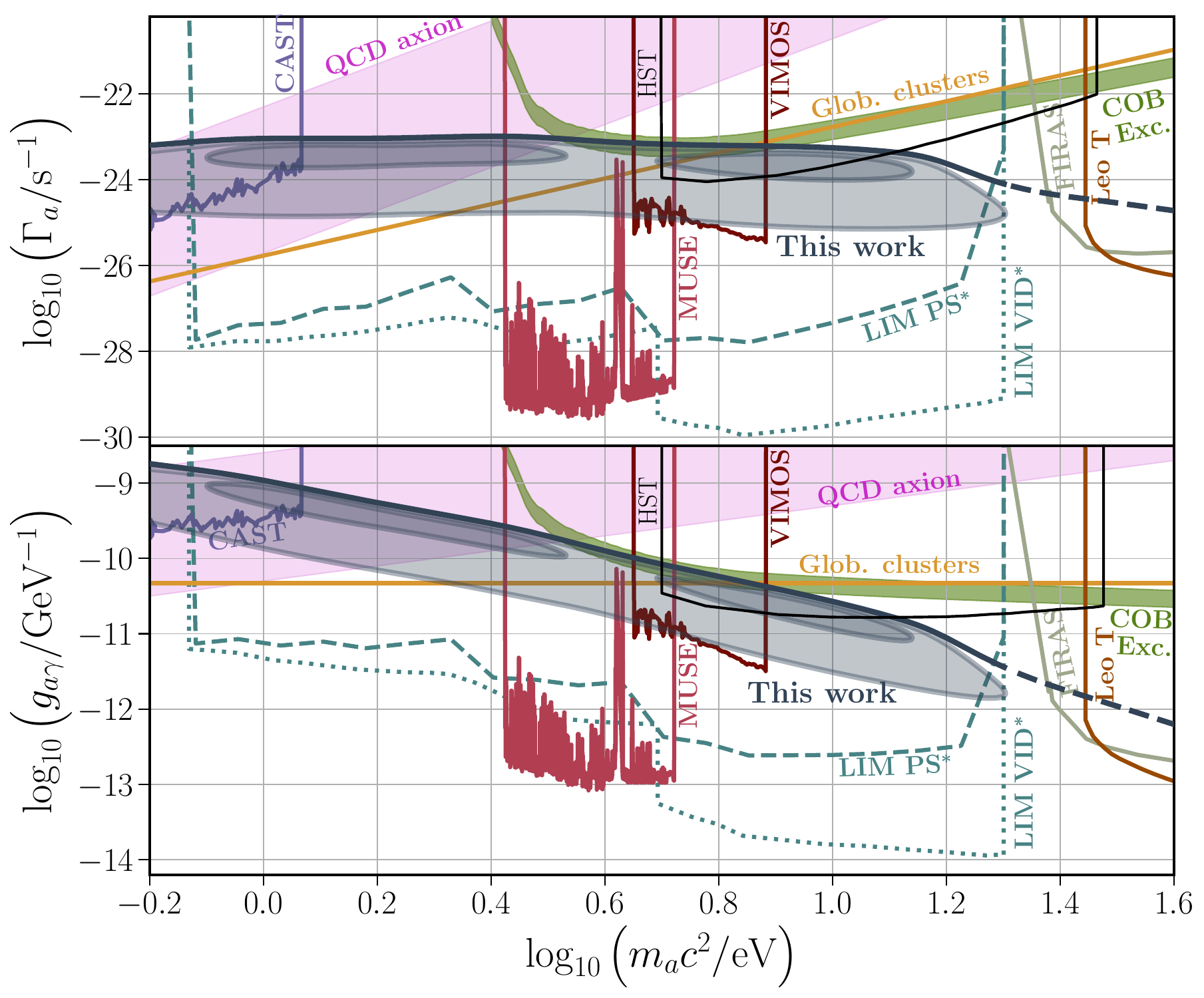}
\caption{68\%, 95\% and 99\% confidence level constraints on the ALP parameters from $\gamma$-ray attenuation (dark blue contours, dashed lines denoting the masses that are affected by neutral hydrogen absorption), along with current strongest 95\% confidence level bounds and preferred region to explain the COB excess, and the QCD axion band.}  
\label{fig:constraints}
\end{centering}
\end{figure}

We compare our results for the ALP decay rate with the current strongest bounds over the relevant mass range in Fig.~\ref{fig:constraints}. We include bounds from the helioscope CAST~\cite{CAST},  from spectroscopic observations of the galaxy clusters Abell 2667 and 2390 with VIMOS~\cite{Grin:2006aw} and the dwarf spheroidal galaxy Leo T with MUSE~\cite{Regis:2020fhw}, the study of the stellar cooling in globular clusters~\cite{Ayala:2014pea, Dolan:2022kul}, cosmic microwave background spectral distortions measured with COBE/FIRAS~\cite{Bolliet_SD},  radiative gas cooling rates of Leo T~\cite{Wadekar_LeoTDMheating}, and HST angular power spectrum~\cite{Nakayama:2022jza};  and forecast sensitivities of line-intensity mapping~\cite{Bernal_limdm}.\footnote{We include forecasts using the voxel intensity distribution (VID, an estimator for the probability distribution function of the line intensity within a voxel), and the power spectrum (PS).} We do not include outdated constraints from observations of the optical and near-infrared background~\cite{Cadamuro:2011fd} and instead add the preferred parameter values to explain the cosmic optical background excess~\cite{Bernal_COB}. The bimodal distribution is produced by the difficulty to fit the local and high $z_{\rm s}$ measurements at the same time (see App.~\ref{sec:AppC}). 

The fact that the optical depth excess may also be explained by additional EBL from galaxies at $z<6$ may point to a scenario where both additional components contribute to the EBL. The exploration of this hypothesis requires a more detailed modeling for the additional astrophysical EBL and is left for future work. Therefore, the hint found in this work shall be confirmed by subsequent studies and additional, independent probes of ALP, and we prefer to interpret this result as an upper bound in the ALP-photon coupling and a motivation for further exploration of this region of the parameter space.

\section{Conclusions}
\label{sec:conclusions}
$\gamma$-ray attenuation returns the strongest $3\sigma$ bound to date on the effective ALP-photon coupling on the $8-25$ eV mass range, improving current bounds by up to more than an order of magnitude. These results rule out most of the parameter space that may explain the cosmic optical background excess, leaving a small window at $m_ac^2\sim 8$ eV. However, note that Ref.~\cite{Bernal_COB} does not include contributions from IHL and galaxies at $z>6$, which may push the viable parameter combinations to lower effective couplings, increasing the compatibility with this work. 

Our analysis is significantly more sensitive to ALP contributions than the study from Ref.~\cite{Korochkin_axion}, based on fitting a bump feature in the spectrum of a single blazar. This is because we use optical-depth measurements from almost 800 blazars, making our study more robust statistically and regarding uncertainties in the intrinsic spectra of the blazars. We rule out most of the viable favored parameter space reported in Ref.~\cite{Korochkin_axion} and we find potential preference for lower couplings.

$\gamma$-ray attenuation as a probe for ALPs or other non-standard contributions to the EBL has great potential to improve in the near future. Existing $\gamma$-ray satellites and ground Cherenkov telescopes like Fermi-LAT and HAWC\footnote{\url{https://www.hawc-observatory.org/}}, HESS\footnote{\url{https://www.mpi-hd.mpg.de/hfm/HESS/}}, MAGIC\footnote{\url{https://magic.mpp.mpg.de/}} or VERITAS\footnote{\url{https://veritas.sao.arizona.edu/}} will keep improving the measurements and increasing the sensitivity to the EBL, while forthcoming facilities such as CTA~\cite{CTAConsortium:2017dvg} will bring dramatic improvements. On the other hand, improved observations of the EBL such as a reassessment of HST data~\cite{Windhorst:2022oip} or observations with SPHEREx~\cite{SPHEREx} and JWST\footnote{\url{ https://www.jwst.nasa.gov/}}, will reduce uncertainties in the EBL determination and increase the precision and robustness of analyses like this one. Furthermore, line-intensity mapping observations will distinguish between different potential sources for exotic lines~\cite{Bernal_limdm, Bernal:2021ylz}, while a non detection would imply that any additional contribution must have a broad spectrum, increasing the characterization of any non-accounted for contributions to the EBL and $\gamma$-ray attenuation.  

%%%%%%%%%%%%%%%%%%%%%%%%%%%%%
\acknowledgments%
%%%%%%%%%%%%%%%%%%%%%%%%%%%%%
JLB is supported by the Allan C.\ and Dorothy H. Davis Fellowship. AC is supported by the Foreign Postdoctoral Fellowship Program of the Israel Academy of Sciences and Humanities and also acknowledges support from the Israel Science Foundation (Grant 1302/19) and the European Research Council (ERC) under the EU Horizon 2020 Programme (ERC-CoG-2015-Proposal n. 682676 LDMThExp). GSP was supported by the National Science Foundation Graduate Research Fellowship under Grant No.\ DGE1746891.  MK was supported by NSF Grant No.\ 2112699 and the Simons Foundation.

%%%%%%%%%%%%%%%%%%
%%%%%%%%%%%%%%%%%%

\appendix

\section{Astrophysical EBL model}
\label{sec:AppA}
Here we provide more details on the theoretical models used to determine the contributions to the EBL from galaxies at $z>6$ and the IHL.

\subsection{Galaxies at $z>6$}
\label{sec:ares}
Our model for the contribution to the EBL from galaxies at $z>6$ has two components. In order to be conservative, we adopt two extreme cases, and take the envelope of the resulting minimum and maximum EBL as our lower and higher value uncertainties.

First, we include an empirically-calibrated model for Pop~II star-forming galaxies \cite{Mirocha2017}, which relates the star-formation rate to the accretion rate of dark matter halos, and infers the star formation efficiency through fits to high-$z$ galaxy luminosity functions from \cite{Bouwens2015}. We use the \textsc{BPASS} version 1.0 single-star models \cite{Eldridge2009} to generate our model galaxy spectrum, assuming solar metallicity, no attenuation from dust, and that stars have been forming at a constant rate for 100 Myr. This is a reasonable approximation for the rest-ultraviolet portion of high-$z$ galaxy spectra, which is the most relevant for the $\gamma$-ray opacity from the EBL. 

We also include a contribution from Pop~III stars, which form in low-mass ``minihalos'' with virial temperatures below the atomic cooling threshold. We either do not account for contributions from Pop III stars or assume that Pop~III stars dominate the cosmic star-formation rate density until $z \simeq 15$ and are extinct by $z \simeq 10$. This scenario is achieved assuming a simple parameterization~\cite{Mirocha:2017xxz}, in which each halo forms Pop~III stars for 250 Myr before transitioning to Pop~II star formation. Pop~III stars are assumed to be massive, $\gtrsim 100 \ M_{\odot}$ stars with hard spectra, comparable to a black body at $10^5$ K~\cite{Schaerer2003}. Such a model will be testable with forthcoming observations by SPHEREx~\cite{Sun2021}.

In addition, we assume that each stellar source population described above is also accompanied by high-mass X-ray binaries, which contribute to the X-ray background. For these sources, we assume a multi-colour disk spectrum~\cite{Mitsuda1984}, with no intrinsic attenuation. We consider cases in which the relationship between X-ray emission and the star-formation rate is similar to the local relation~\cite{Mineo2012}, as well as a case where it is boosted by a factor $10^3$ with respect to the local relation. In both cases, the contribution to the opacity from the EBL is negligible, since the range of energies of the redshifted X-ray photons do not annihilate $\gamma$-ray photons efficiently.

We model these high-$z$ sources and their contribution to the EBL using the publicly-available \textsc{ARES} code~\cite{Mirocha:2014faa}. Note that the lowest energy of the $z>6$ contribution is set by the lowest energy considered within \textsc{ARES}, which is the minimum photon included in the BPASS library $\simeq 0.4$ eV. Nevertheless, this energy is always below $\epsilon'_{\rm min}$, hence not affecting the results. 

%%%%%%%%%%%%%%%%%%%%%%%%%%%%%%%%%%%%%%%%%%%%%%%%%%%%%%%%%%%%%%%%%%%%%%%%%%%%%%%
\subsection{Intra-halo light} 
%%%%%%%%%%%%%%%%%%%%%%%%%%%%%%%%%%%%%%%%%%%%%%%%%%%%%%%%%%%%%%%%%%%%%%%%%%%%%%%

Diffuse intra-halo emission, usually referred to as intra-halo light (IHL), may come from tidally stripped stars during galaxy mergers, with aggregated surface brightness low enough to challenge resolved observations~\cite{Conroy:2007cv}. The fraction of stripped stellar mass depends on the halo mass, with heavier halos expected to host larger fractions of diffuse intra-halo emission than their lighter counterparts.  We follow the parametric model from Refs.~\cite{Mitchell-Wynne:2015rha, Cooray12}. The specific luminosity density at redshift z and rest-frame wavelength $\lambda$ can be computed using the halo mass function $\dif n_{\rm h}/\dif M$ as
\begin{equation}
    j_{\lambda,{{\rm IHL}}}(\lambda,z) = \int_{M_{\rm min}}^{M_{\rm max}} \dif M \frac{\dif n_{\rm h}}{\dif M}L_{\lambda,{\rm IHL}}(M,z),
\end{equation}
where we use the halo mass function from Ref.~\cite{Tinker:2008ff}, we assume $M_{\rm min}=10^9\, M_\odot/h$ and $M_{\rm max}=10^{13}\,M_\odot/h$, where $h=H_0/(100\,{\rm km/s/Mpc})$ in this context, and $L_{\lambda,{\rm IHL}}$ is the intra-halo light specific luminosity emitted by a halo of mass $M$, given by
\begin{equation}
    L_{\lambda,{\rm IHL}}(M,z)=f_{\rm IHL}(M)L_{2.2}(M)(1+z)^{\alpha}S_{\lambda, 2.2}(\lambda),
    \label{Eq:LumiMassRelation}
\end{equation}
where $L_{2.2}\equiv L_0/(2.2\,\mu{\rm m})$ is the total halo luminosity at 2.2 $\mu$m at $z=0$, with $L_0$ given by~\cite{Lin:2004ak}
\begin{equation}
 L_0 = 5.64\times 10^{12} h_{70}^{-2} \left(\frac{M}{2.7 \times 10^{14}h_{70}^{-1}M_{\odot}}\right)^{0.72}L_{\odot}\,,
\end{equation}
where $h_{70}=H_0/(70\,{\rm km/s/Mpc})$. In Eq.~\eqref{Eq:LumiMassRelation}, $\alpha$ controls the redshift evolution of the luminosity and $f_{\rm IHL}$ is the fraction of the total halo luminosity coming from IHL. We parametrize $f_{\rm IHL}$ with a power law in mass
\begin{equation}
    f_{\rm IHL} = A_{\rm IHL} \left(\frac{M}{10^{12}M_{\odot}}\right)^{\beta}\,.
\end{equation}
Finally, $S_{\lambda,2.2}\equiv S_\lambda/S_{2.2}$ is the spectral energy distribution of the IHL normalized to be 1 at 2.2 $\mu$m. We assume $S_\lambda$ to be similar to old elliptical galaxies, comprised of red stars~\cite{Krick:2007fp}. In practice, we use the template for an elliptical galaxy of age 13 Gyr from the \textit{SWIRE} library~\cite{Polletta:2007ut}. Once we have computed $j_\lambda$, we transform it to $j_\nu$ and compute the contributions to the EBL using Eq.~\eqref{eq:lamIlam}.

We follow the results of Ref.~\cite{Mitchell-Wynne:2015rha}, which assumes a fixed index for the mass power law to 0.1 (in agreement with the best fit to the near-infrared background power spectrum, $\beta = 0.094 \pm 0.005$~\cite{Cooray_2012}), and finds the index of the redshift power law and the $\log_{10}A_{\rm IHL}$ completely anticorrelated, and fairly uncorrelated with other contributions. We set $\log_{\rm 10} A_{\rm IHL} =  \left\lbrace -3.35,\, -3.23,\, -3.09 \right\rbrace$ and $\alpha = \left\lbrace 0.1,\, 1,\, 1.5 \right\rbrace$ as the low, mean and high values for our estimation, according to the reported 68\% confidence level constraints in Ref.~\cite{Mitchell-Wynne:2015rha}.

%%%%%%%%%%%%%%%%%%%%%%%%%%%%%%%%%%
%%%%%%%%%%%%%%%%%%%%%%%%%%%%%%%%%%

\begin{figure*}[t]
 \begin{centering}
\includegraphics[width=\textwidth]{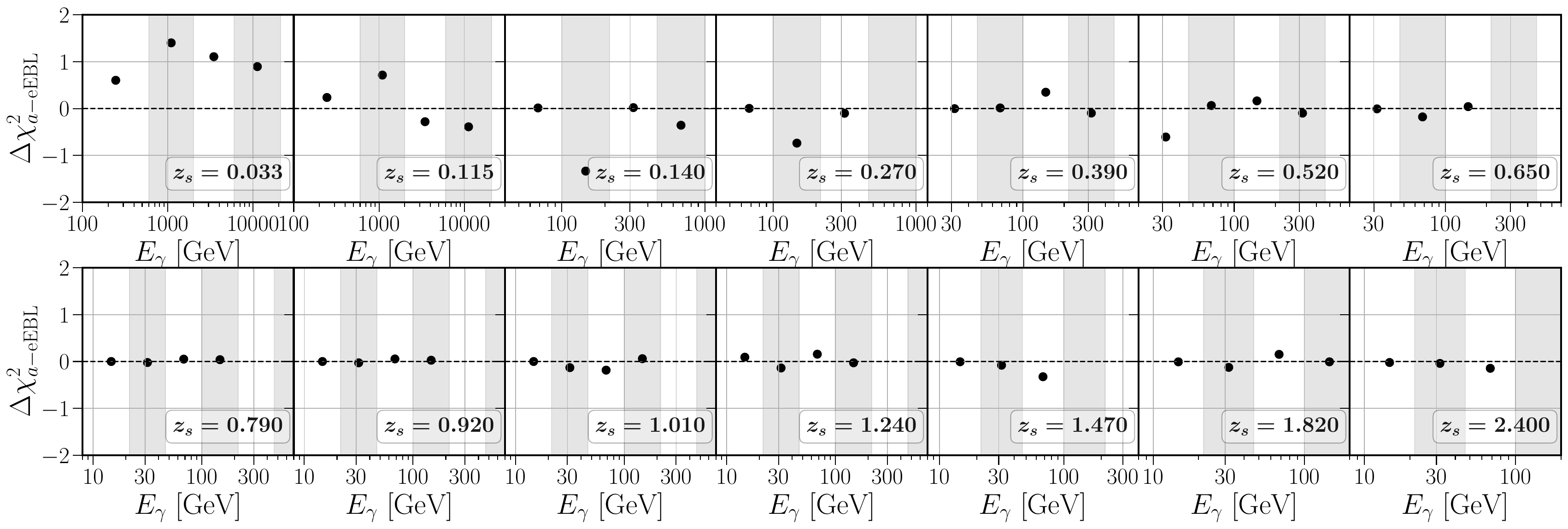}
\caption{Comparison of the individual $\chi^2$ values for the best-fit cases in the ALP and $F_{\rm eEBL}$ scenarios.}  
\label{fig:Dchi2_individual}
\end{centering}
\end{figure*}

\section{Likelihood}
\label{sec:AppB}
In this appendix we provide further detail about the likelihood we use for the residual optical depth $\tau_{\rm res}$. Let us first describe how we obtain  $\tau_{\rm res}$ for each source redshift $z_{\rm s}$ and observed energy $E_\gamma$ from the measured optical depth $\tau_{\rm meas}$ and the predicted contributions to the optical depth $\tau_{k}$ from each component of the astrophysical EBL (gal. $z<6$, gal. $z>6$, and IHL), indexed by $k$. We index the $z_{\rm s}$ and $E_\gamma$ values with $i$ and $j$, respectively, so that $\tau(z_{\rm s}^{(i)},E_\gamma^{(j)})\equiv \tau^{(ij)}$. As discussed in the main body of the article, we discard lower-limit measurements of $\tau_{\rm meas}$,  and distinguish between actual measurements and only upper limits for $\tau_{\rm meas}$. If $\tau_{\rm meas}^{(ij)}$ is detected, we compute $\tau_{\rm res}^{(ij)}$ as
\begin{equation}
    \tau_{\rm res}^{(ij)}=\tau_{\rm meas}^{(ij)}-\sum_k \tau_{k}^{(ij)}\,,
\end{equation}
with associated 68\% confidence level uncertainties obtained adding in quadrature the measurement errors and the uncertainties from the astrophysical contributions:
\begin{equation}
    \sigma_x^2(\tau_{\rm res}^{(ij)})=\sigma_x^2(\tau_{\rm meas}^{(ij)}) + \sum_k \sigma_x^2(\tau_{k}^{(ij)})\,,
\end{equation}
where the subscript $x$ denotes whether these are upper- or lower-value errors ---$\sigma_{\rm up}$ and $\sigma_{\rm low}$, respectively. We consider the cases for which $\tau_{\rm res}^{(ij)}<\sigma_{\rm low}(\tau_{\rm res}^{(ij)})$ (i.e., the low end of the 68\% confidence level uncertainties lies below zero) as upper limits, set by
\begin{equation}
    \sigma_{\rm up\, lim}(\tau_{\rm res}^{(ij)}) = \left({\rm CDF}_{(ij)}\right)^{-1}(0.683)\,,
\end{equation}
which is the inverse of the cumulative distribution function of a normalized asymmetric Gaussian with only $\tau>0$ and standard deviations given by $\sigma_{\rm up}(\tau_{\rm res}^{(ij)})$ and $\sigma_{\rm low}(\tau_{\rm res}^{(ij)})$, and consider $\tau_{\rm res}^{(ij)}=0$.

Finally, we include the measured upper limits for the optical depth in our analysis, since they set a maximum to the contributions from unknown components of the EBL. In these cases, we consider $\tau_{\rm res}^{(ij)}=0$, and the 68\% confidence level upper limit to be
\begin{equation}
\begin{split}
    \sigma_{\rm up\, lim}^2(\tau_{\rm res}^{(ij)}) & = \left(\sigma_{\rm up\, lim}(\tau_{\rm meas}^{(ij)})-\sum_k \tau_k^{(ij)}\right)^2 +\\
    & + \sum_k \sigma_{\rm up}^2(\tau_{k}^{(ij)})\,.
\end{split}
\end{equation}
We subtract $\tau_k$ to the measured upper limit to take into account the fact that there is a \textit{known} contribution below the detection limit. Therefore, additional contributions to the optical depth have to be added to that, after adding the astrophysical uncertainties in quadrature to the measured upper limit.

Once that we have the computed $\tau_{\rm res}^{(ij)}$ and its corresponding uncertainties, we can build the likelihood for additional contributions to the optical depth, which we denote with the subscript `extra'.  The value of $\tau_{\rm extra}$ depends on the parameters considered in each model, which we denote here with $\Theta$. Following Ref.~\cite{Fermilat_EBL18}, we assume a Gaussian likelihood $\mathcal{L}\propto \exp(-\chi^2/2)$, considering that all measurements are uncorrelated, such as
\begin{equation}
    \chi^2=
    \begin{cases}
    \sum_{ij} \frac{\left(\tau_{\rm extra}^{(ij)}(\Theta)-\tau_{\rm res}^{(ij)}\right)^2}{\sigma_{\rm up}^2(\tau_{\rm res}^{(ij)})} & {\rm if \, \tau_{\rm extra}^{(ij)}(\Theta)\geq\tau_{\rm res}^{(ij)}}\,,\\
    \sum_{ij} \frac{\left(\tau_{\rm extra}^{(ij)}(\Theta)-\tau_{\rm res}^{(ij)}\right)^2}{\sigma_{\rm low}^2(\tau_{\rm res}^{(ij)})} & {\rm if \tau_{\rm extra}^{(ij)}(\Theta)<\tau_{\rm res}^{(ij)}}\,,\\
    \sum_{ij} \frac{\left(\tau_{\rm extra}^{(ij)}(\Theta)\right)^2}{\sigma_{\rm up\, lim}^2(\tau_{\rm res}^{(ij)})} & {\rm if \tau_{\rm res}^{(ij)}=0}\,.\\ 
    \end{cases}
    \label{eq:chi2}
\end{equation}
Given that the maximum number of parameters that we consider in our models is two, we use a fine grid in parameter space to evaluate Eq.~\eqref{eq:chi2}, and obtain the unnormalized posterior distribution from $\Delta \chi^2(\Theta)\equiv \chi^2(\Theta)-\chi^2(\Theta_{\rm bf})$, where $\Theta_{\rm bf}$ are the best-fit parameter values.

%%%%%%%%%%%%%%%%%%%%%%%%%%%%%%%%%%
%%%%%%%%%%%%%%%%%%%%%%%%%%%%%%%%%%

\section{Null tests}
\label{sec:AppC}
Here we provide further comparison between the ALP case (the main case of study considered in this work) and two null tests. We focus on the EBL from galaxies at $z<6$ since it is by far the dominant contribution. We consider \textit{a}) an overall boost of the astrophysical EBL from galaxies at $z<6$, multiplying the standard contribution by $(1+F_{\rm eEBL})$ factor; and \textit{b}) a potential underestimation of the EBL uncertainties, boosting the uncertainties of the contribution from galaxies at $z<6$ by a constant factor.

\subsection{ALPs vs $F_{\rm eEBL}$}
We have shown in the main text that an overall $14-30\%$ increase of the EBL can explain the optical depth excess with similar preference over the null hypothesis than ALP decays. Here we extend the comparison between the two hypotheses. In order to compare how well each case fit $\tau_{\rm res}$ we compare their $\chi^2$ values for their best fit, defining a $\Delta\chi^2_{a-{\rm eEBL}}\equiv\chi^2_{a,{\rm bf}}-\chi^2_{\rm eEBL, bf}$; $\Delta\chi^2_{a-{\rm eEBL}}>0$ denotes preference for the $F_{\rm eEBL}$ hypothesis and viceversa.  

The results shown in Fig.~\ref{fig:chi2} seem to indicate that $F_{\rm eEBL}$ and ALP decays fit better the more local and more distant blazars, respectively. We confirm this in Fig.~\ref{fig:Dchi2_individual}, where we show $\Delta\chi^2_{a-{\rm eEBL}}$ for each bin in $z_{\rm s}$ and $E_{\gamma}$. While for most cases the two models perform similarly, there is a clear preference for ALPs at intermediate redshifts, while measurements at low redshifts and energies favor $F_{\rm eEBL}$. More quantitatively, we find  $\Delta\chi^2_{a-{\rm eEBL}} ) = 0.79$ when considering all cases, which shifts to $-2.9$ if the first $z_{\rm s}$ bin is not considered in the whole analysis.

\begin{figure}[t]
 \begin{centering}
\includegraphics[width=\columnwidth]{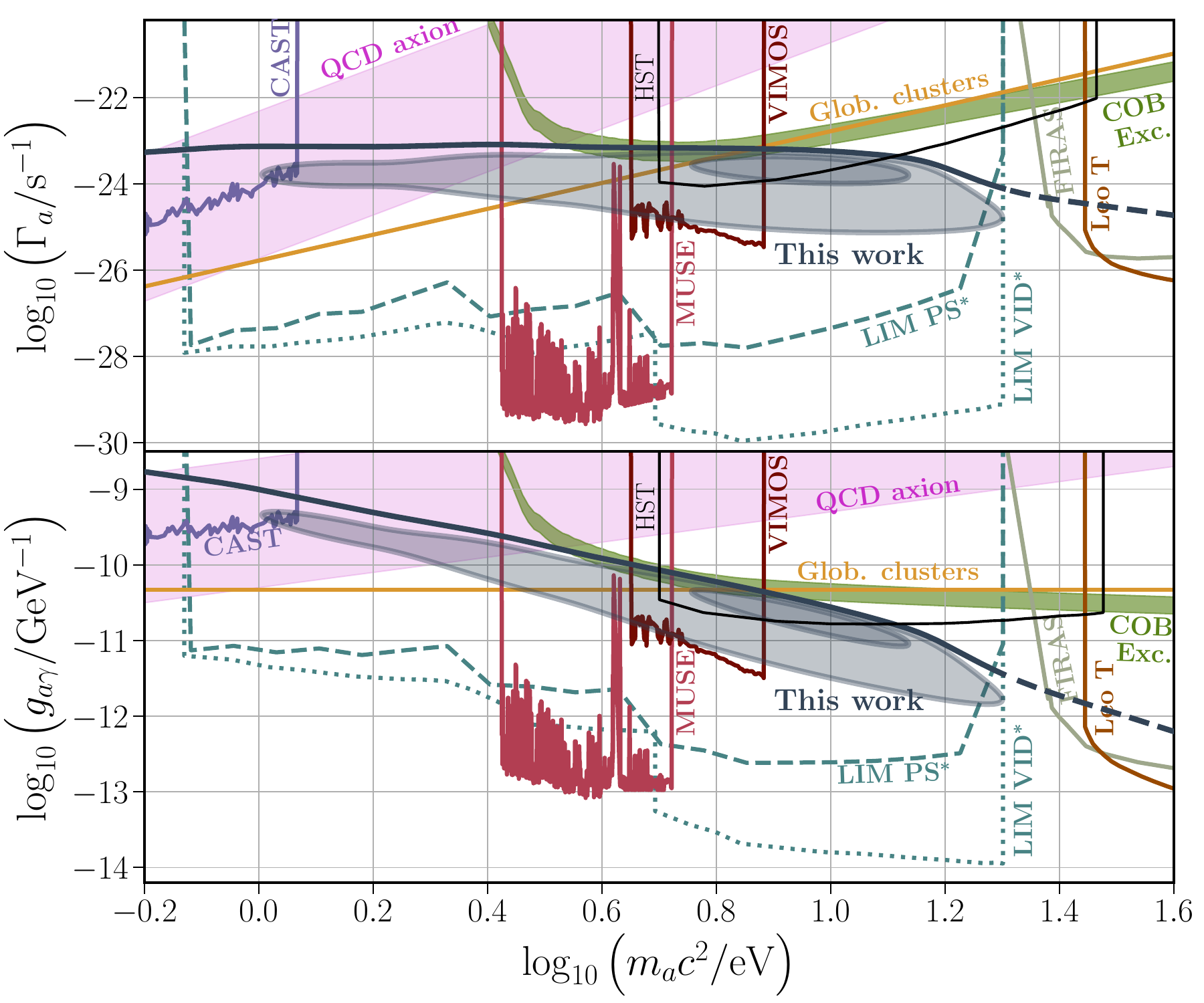}
\caption{Same as Fig.~\ref{fig:constraints} but removing the first $z_{\rm s}$ bin from the analysis.} 
\label{fig:no_zs1}
\end{centering}
\end{figure}

In Fig.~\ref{fig:no_zs1} we show how the peak in the posterior at light masses disappears when the first $z_{\rm s}$ bin is removed. This proves that that peak was present only due to the first redshift bin (which, as shown in Fig.~\ref{fig:chi2}, does not fit satisfactorily. In this case, the best fit $F_{\rm eEBL}=0.17$, is preferred over the null hypothesis at $1.9\sigma$ significance ($p-{\rm value}=5.6\times 10^{-2}$). On the other hand, the best fit and significance for the ALP-decay case remains very similar. These results further support the scenario in which, if there are actually contributions from ALP decays,  they may be accompanied by additional astrophysical EBL to accomplish a good fit to both local and distant blazars.

\begin{figure}[h]
 \begin{centering}
\includegraphics[width=\columnwidth]{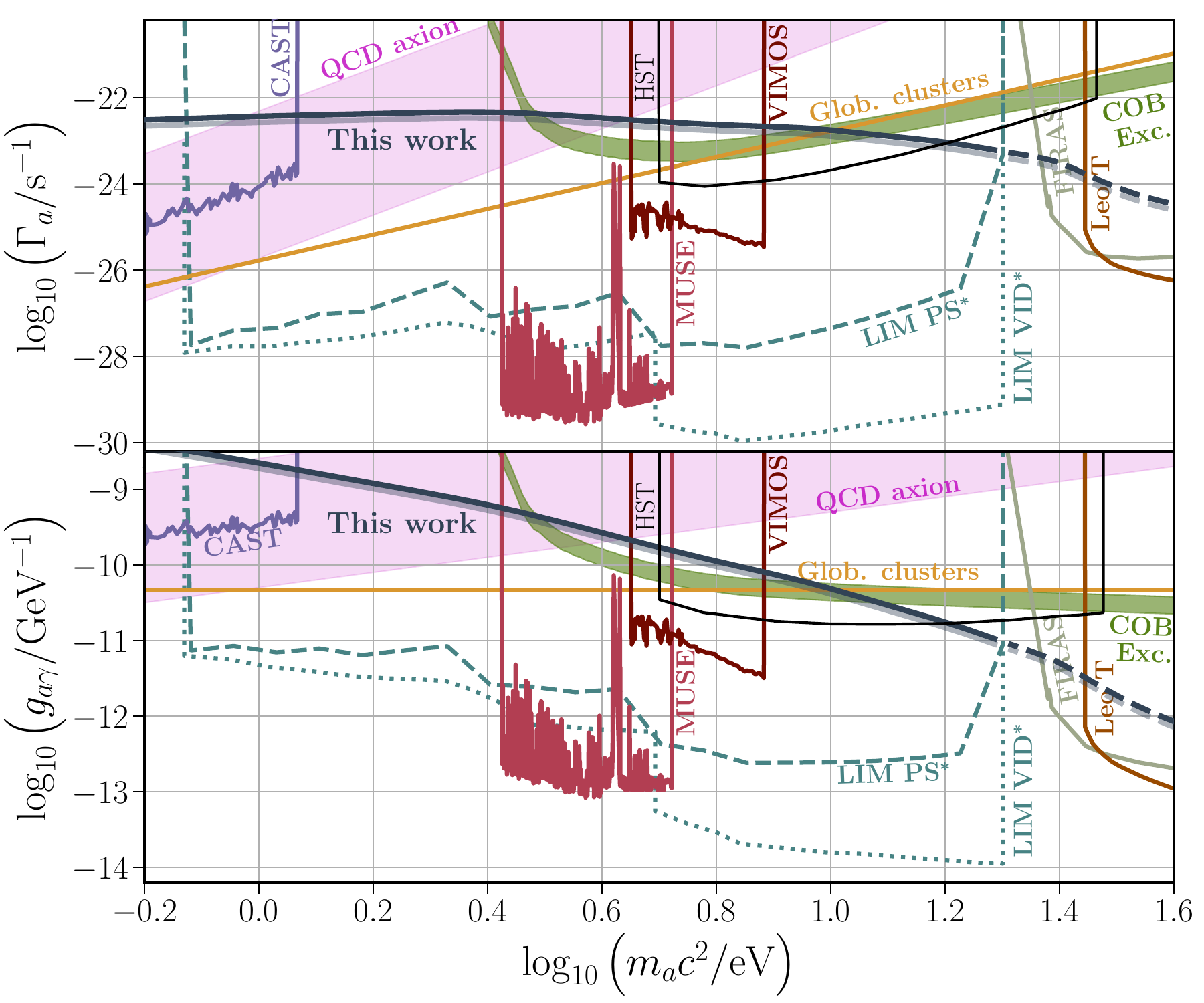}
\caption{95\% and 99\% confidence level bounds on the ALP-photon effective coupling as function of mass from $\gamma$-ray attenuation for the extremely conservative scenario discussed in App.~\ref{sec:boosted} (dark blue lines, dashed lines denoting the masses that are affected by neutral hydrogen absorption), along with current strongest 95\% confidence level bounds and preferred region to explain the COB excess, and the QCD axion band.}  
\label{fig:constraints_boost}
\end{centering}
\end{figure}

\subsection{Increasing EBL uncertainties}
\label{sec:boosted}
We carry out a final null test. In this case, we boost the uncertainties of the astrophysical EBL from galaxies at $z<6$ by a constant factor until $\tau_{\rm res}$ becomes an upper limit for all $z_{\rm s}$ and $E_{\gamma}$ bins considered in the analysis. We find that the required factor is 14.36. By construction, in this case we do not find any preference for contributions from ALPs over the null hypothesis. However, we can set extremely conservative bounds on the ALP-photon effective coupling. We show the corresponding 95\% and 99\% confidence level bounds in Fig.~\ref{fig:constraints_boost}. Even under these extremely conservative assumptions we find that the $\gamma$-ray attenuation significantly improves over current upper bounds by up to an order of magnitude. This result further motivates the study of $\gamma$-ray attenuation as a very promising probe of ALP and other exotic contributions to the EBL.

\bibliography{Refs.bib}
\bibliographystyle{utcaps}

\end{document}